\documentclass{aastex6}
\usepackage{epsfig}
\usepackage{graphicx}
\usepackage{float}
\usepackage{amsmath}
\usepackage{color}
\usepackage{amssymb}
\usepackage{amsfonts}
\usepackage{units}
\usepackage{bm}
\usepackage{lineno}
\usepackage{epstopdf}
\usepackage{amstext}

\begin{document}

\title{Afterglow polarizations in a stratified medium with effect of the equal arrival time surface}

\author{Mi-Xiang Lan$^{1}$, Xue-Feng Wu$^{2,3}$, and Zi-Gao Dai$^{4}$}
\affil{$^{1}$Center for Theoretical Physics and College of Physics, Jilin University, Changchun, 130012, China; lanmixiang@jlu.edu.cn \\
$^{2}$ Purple Mountain Observatory, Chinese Academy of Sciences, Nanjing 210023, China \\
$^{3}$School of Astronomy and Space Sciences, University of Science and Technology of China, Hefei 230026, China\\
$^{4}$ Department of Astronomy, School of Physical Sciences, University of Science and Technology of China, Hefei 230026, China\\}

\begin{abstract}
The environment of gamma-ray burst (GRB) has an important influence on the evolution of jet dynamics and of its afterglow. Here we investigate the afterglow polarizations in a stratified medium with the equal arrival time surface (EATS) effect. Polarizations of multi-band afterglows are predicted. The effects of the parameters of the stratified medium on the afterglow polarizations are also investigated. We found the influences of the EATS effect on the afterglow polarizations become important for off-axis detections and PD bumps move to later times with the EATS effect.  Even the magnetic field configurations, jet structure and observational angles are fixed, polarization properties of the jet emission could still evolve. Here, we assume a large-scale ordered magnetic field in the reverse-shock region and a two-dimensional random field in the forward-shock region. Then PD evolution is mainly determined by the evolution of $f_{32}$ parameter (the flux ratio between the reverse-shock region and forward-shock region) at early stage and by the evolution of the bulk Lorentz factor $\gamma$ at late stage. Through the influences on the $f_{32}$ or $\gamma$, the observational energy band, observational angles, and the parameters of the stratified medium will finally affect the afterglow polarizations.
\end{abstract}

\keywords{Gamma-ray bursts (629); magnetic fields (994);}

\section{Introduction}
Short-duration gamma-ray bursts (sGRBs) are proved to be originated from the mergers of the double compact objects \citep{Abbott2017}, while long-duration gamma-ray bursts (lGRB) are associated with collapse of massive stars \citep{Mazzali2003}. So the environments of sGRBs would be the interstellar medium (ISM), while they could be the stellar wind for lGRBs. However, the works \citep{Yi2013} had suggested that the environments of GRBs would be neither a uniform ISM with a constant density nor a stellar wind with a number density $n(r)$ proportional to $r^{-2}$. Assuming a power-law profile of the environment density, i.e., $n(r)\propto r^{-k}$, it is found that the typical value of $k$ is $\sim1$ \citep{Yi2013}.

Since there might be large-scale ordered magnetic field in the reverse-shock region, carried out with the outflow from the central engine, the emission from the reverse-shock region would be highly polarized. The profiles of the GRB environments would affect the dynamics and the emission during afterglow phase. Besides the forward shock emission, the contribution of emission from the reverse-shock region to the total jet emission would also be affected by the environment \citep{Kobayashi2000,CL2000,Wu2003}. Therefore, the environment may affect the polarizations of early afterglow. Because of relativistic motion, radiations at different radius $r$ will arrive the observer at same observational time $t$ and the locus of these radii forms the equal arrival time surface (EATS, \cite{Sari1998}). EATS effect would become important for a stratified medium. Therefore, afterglow polarizations with EATS effect should be investigated.

Afterglow polarization had been investigated widely in the literature. Polarization of the jet emission in a three-dimensional anisotropic random magnetic field were considered by \cite{Sari1999} and \cite{Gruzinov1999}. The afterglow polarizations with various jet structure were studied by \cite{Rossi2004}, \cite{Lazzati2004}, \cite{Wu2005}, and \cite{Lan2018}. \cite{GK2003} discussed the afterglow polarizations with an large-scale ordered magnetic field component in the ambient medium. \cite{Lazzati2004} had also discussed the late-time afterglow polarization with a toroidal magnetic field component in the shocked ISM. Afterglow polarizations considering both the reverse-shock and the forward-shock emission was investigated by \cite{Lan2016}. Recently, afterglow polarizations of an off-axis top-hat jet with lateral expansion in a stratified medium was discussed \citep{Pedreira2022}. However, they did not include the EATS effect in their treatment which might be important for off-axis detections \citep{Huang2007}.

In this paper, we consider afterglow polarizations with EATS effect in an arbitrary outer medium with a power-law number density. And two cases are studied, i.e., the reverse-shock emission dominates the early afterglow (Case I) and the forward-shock radiation dominates the emission of the whole afterglow phase (Case II). The paper is arranged as follows. In Section 2, we describe our model. Our new results with EATS effect and other revises, compared with the results in \cite{Lan2016}, are shown in Section 3. Afterglow polarizations in a stratified medium are presented in Section 4. Finally, our conclusions and discussion are in Section 5. Throughout the paper, a flat universe with $\Omega_\Lambda=0.73$, $\Omega_M=0.27$, and $H_0=71$ km s$^{-1}$ Mpc$^{-1}$ is adopted.

\section{The model}

\subsection{The dynamics}
An ultrarelativistic outflow from a GRB central engine is usually thought to be collimated. With the propagation of this collimated outflow ( i.e. the jet) in an outer medium, two shocks (i.e., the reverse and forward shocks) will be formed. So there are four regions in system: the unshocked outflow (Region 4), the shocked outflow (Region 3 or the reverse-shocked region), the shocked medium (Region 2 or the forward-shocked region), and the unshocked medium (Region 1). Observations of early optical flashes suggest the magnetization of GRB outflow could not be high ($\sigma\leq1$) \citep{ZK2005,Lan2016}. Therefore, the magnetic field will be dynamically unimportant. The dynamic model here follows that in \cite{Lan2016}, where the energy conservation of the system is used to derive the dynamics. A homogeneous interstellar medium (ISM) was considered in \cite{Lan2016}, here we generate our study to a stratified medium with a power-law density distribution of $n(r)=n_0(r/r_0)^{-k}$. $r$ is the radius from central engine. We fix $n_0=1\ cm^{-3}$ throughout this paper and consider the effects of different values of $r_0$ and $k$ on the dynamics. The lateral expansion is not considered and the half-opening angle of the top-hat jet is fixed to be $\theta_j=0.1$ rad.

With the EATS effect, it is convenient to express the dynamical quantities (e.g., the bulk Lorentz factor $\gamma$) as functions of the radius $r$. The EATS used here reads \citep{Sari1998}
\begin{equation}
t_b-\frac{r\cos\theta}{c}=\frac{t}{1+z}
\end{equation}
where $t_b$ is the dynamic time at the burst source frame. $\theta$ is the angle between the local velocity direction and the line of sight. $c$ is the speed of the light and $z$ is the redshift of the source. We assume before the initial emission radius $r_b$ the outflow moves with a constant initial Lorentz factor $\eta$. So we have $t_b(r_b)=r_b/\beta_0c$, where the initial dimensionless velocity is $\beta_0=\sqrt{1-\eta^{-2}}$. In this paper, we set $r_b=10^{13}$ cm.

\subsection{Afterglow polarizations}
High-level PDs observed in GRB afterglow phase \citep{Steele2009,Mundell2013} indicate that there would be large-scale magnetic field remnants in the afterglow emission region. Theoretically, large-scale ordered magnetic field can be carried with the outflow from the GRB central engine \citep{BZ1977,Spruit2001,Drenkhahn2002}. Because of conservation of magnetic flux, the radial and transverse components of the ordered magnetic field will decrease as $r^{-2}$ and $r^{-1}$, respectively. At large radii (e.g., the emission region of an afterglow), the transverse component will dominate. Literally, there are two kinds of transverse field \citep{Spruit2001,Drenkhahn2002}. One is toroidal configuration, corresponding to a parallel rotator (e.g., a black hole). The other is aligned field, usually related to an inclined central engine (e.g. a magnetar). Here, the radiation mechanism in the afterglow phase is assumed to be synchrotron emission and the inverse-Compton scattering is not considered. As in \cite{Lan2016}, we assume the magnetic field in Region 3 is large-scale ordered and neglect the random field. So our results of PD will be the upper limits for the reverse-shock-emission dominated cases at early stage. In the forward shock region, a two-dimensional random field confined in the shock plane is assumed, which is likely to be generated or amplified by the forward shock. Therefore, the emission from reverse-shock region with large-scale ordered magnetic field will be highly polarized, while the emission from forward-shock region with a two-dimensional random field will be lowly polarized unless it is viewed off-axis \citep{Waxman2003,Lan2016}.

For an aligned field configuration in Region 3 (It orientation is assumed to be $\pi/4$ throughout this paper), the final PDs of the system are always positive ($\Pi_a=\sqrt{Q_\nu^2+U_\nu^2}/f_\nu$) and the polarization directions are depicted by the polarization angles (PAs, $\chi_a=1/2\arctan(U_\nu/Q_\nu)$), where $Q_\nu$, $U_\nu$, and $f_\nu$ are the total Stokes parameter Q, U, and the flux density of the system, including contributions from both Regions 2 and 3 (e.g. $f_\nu=f_{\nu,2}+f_{\nu,3}$, $f_{\nu,2}$ and $f_{\nu,3}$ are the flux density of Regions 2 and 3, respectively.). For a toroidal configuration in Region 3, because of axial symmetry, the Stokes parameter $U_\nu$ of the system is zero. PD of such system is defined as $\Pi_t=Q_\nu/f_\nu$. So depending on the sign of $Q_\nu$, $\Pi_t$ can be positive or negative. And the polarization direction of $\Pi_t>0$ will have a $90^\circ$ difference with that of $\Pi_t<0$. The concrete expressions for the Stokes parameters and the relative quantities all can be found in \cite{Lan2016}. However, the polarization calculation of a two-dimensional random magnetic field in \cite{Lan2016} was not correctly considered, which have been revised in \cite{Lan2019}. Here, the integration will be on the EATS, while it is on the jet surface in \cite{Lan2016}.

In the reverse-shocked or forward-shocked region, the injected energy spectrum of shock-accelerated electrons is $N(\gamma_e)\propto\gamma_e^{-p_i}$. $\gamma_e$ is the Lorentz factor of shock-accelerated electrons. In this paper, we fix the spectral index of the injected electrons in Region $i$ ($i=2$ for Region 2 and $i=3$ for Region 3) to be $p_i=2.5$. The minimum and maximum Lorentz factor of injected electrons in Region $i$ can be expressed as $\gamma_{m,i}=(p_i-2)/(p_i-1)\epsilon_{e,i}e_i/(n_im_ec^2)+1$ and $\gamma_{max,i}=\sqrt{6\pi e/(\sigma_TB'_i)}$, respectively. The internal energy and the number density of Region $i$ are denoted as $e_i$ and $n_i$, respectively. $m_e$ and $e$ are the mass and charge of the electron, respectively. $\sigma_T$ is the cross section of Thomson scattering. The strength of magnetic field in the comoving frame of Region $i$ is denoted as $B'_i$. The cooling of the electrons in the shocked region is also included and the cooling Lorentz factor of electrons is $\gamma_{c,i}=6\pi m_ec/(\sigma_TB^{'2}_it')$. $t'$ is the time in the comoving frame of the shocked region and $t'=\int dt'=\int^r_{r_b}dr/\beta\gamma c$, where $\beta=\sqrt{1-\gamma^{-2}}$ is the dimensionless velocity of the jet.

\section{Comparisons}
\subsection{The Setup}
We compare the results considering EATS effect with that without this effect in \citep{Lan2016}. The parameters of dynamics and emission for the four cases considered here are same as that in \citep{Lan2016}. The fixed dynamical parameters for Cases 1 and 3 are the same (corresponding to the thick shell case): the isotropic equivalent energy $E_{iso}=10^{52}$ erg, initial Lorentz factor $\eta=300$, and initial width of the outflow $\Delta_0=3\times10^{12}$ cm. And the fixed dynamical parameters for Cases 2 and 4 are also the same (corresponding to the thin shell case): $E_{iso}=10^{50}$ erg, $\eta=100$, and $\Delta_0=3\times10^{10}$ cm. The shocks in each case, including both the reverse shock and the forward shock, are assumed to be adiabatic. The redshift of the source is fixed to be $z=1$.

We assume equal-partition of the shocked energy in Regions 2 and 3. A fraction of $\epsilon_{e,i}$ and of $\epsilon_{B,i}$ of the shocked energy go to the electrons and magnetic field, respectively. We fix $\epsilon_{e,3}=\epsilon_{B,3}=0.1$, $\epsilon_{e,2}=0.05$, and $\epsilon_{B,2}=0.002$ in Cases 1 and 4; $\epsilon_{e,3}=0.015$, $\epsilon_{B,3}=0.01$, $\epsilon_{e,2}=0.02$, and $\epsilon_{B,2}=0.005$ in Case 2; $\epsilon_{e,3}=0.01$, $\epsilon_{B,3}=0.005$, $\epsilon_{e,2}=0.02$, and $\epsilon_{B,2}=0.01$ in Case 3. We summarize the parameters of the four cases in the following Table \ref{Table 1}.

\begin{table*}[!htbp]
	\caption{Dynamical and Emission Parameters of the Cases 1, 2, 3, and 4}
	\label{Table 1}
	\begin{center}
		\centering
		\begin{tabular}{ccccccccccc}
			\hline\hline\noalign{\smallskip}
			cases  &  $E_{iso}$&$\eta$&$\Delta_0$&$\epsilon_{e,3}$&$\epsilon_{B,3}$&$\epsilon_{e,2}$&$\epsilon_{B,2}$&\\
			\hline\noalign{\smallskip}
		    Case 1&$10^{52}$ erg&300&$3\times10^{12}$ cm&0.1&0.1&0.05&0.002&\\[5pt]
		    Case 2&$10^{50}$ erg&100&$3\times10^{10}$ cm&0.015&0.01&0.02&0.005&\\[5pt]
            Case 3&$10^{52}$ erg&300&$3\times10^{12}$ cm&0.01&0.005&0.02&0.01&\\[5pt]
            Case 4&$10^{50}$ erg&100&$3\times10^{10}$ cm&0.1&0.1&0.05&0.002&\\[5pt]
            \noalign{\smallskip}
			\hline
		\end{tabular}
	\end{center}
    \footnotesize{\ \ \ \ \ \ \ \ \ \ \ \ \ \ \ \ \ \ \ \ \ The other parameters are the same for four cases: $\theta_j=0.1$ rad, $p_2=p_3=2.5$, $n_1=1$ cm$^{-3}$, and $z=1$.}
\end{table*}

For comparison, as in \cite{Lan2016} an ISM environment ($k=0$) is considered. It should be noted that the sets for repeating the results of \cite{Lan2016} here are the same as that in \cite{Lan2016} except that the polarization calculation for a two-dimensional random magnetic field is corrected. For the new sets here, the differences are that EATS effect is considered, the polarization calculation for a two-dimensional random magnetic field is corrected (see \cite{Lan2019}), and the local PD in an ordered field is expressed as
\begin{equation}
\pi_0=\frac{\int G(x)N(\gamma_e)d\gamma_e}{\int F(x)N(\gamma_e)d\gamma_e},
\end{equation}
where $F(x)=x\int^\infty_xK_{5/3}(t)dt$ and $G(x)=xK_{2/3}(x)$. And $x=\nu'/\nu'_c$, $\nu'=(1+z)\nu/\mathcal{D}$ and $\nu'_c$ are the observational frequency in the comoving frame and the critical frequency of electrons with Lorentz factor $\gamma_e$, respectively. $K_{5/3}(x)$ and $K_{2/3}(x)$ denote the modified Bessel functions with $5/3$ and $2/3$ orders. It should be noted that in the forward-shock-dominated cases (i.e., Cases 2 and 3) the emission from the reverse shock region is also included although it is unimportant for the whole jet emission.

\subsection{The results}
\begin{figure*}
	\centering \includegraphics[width=\textwidth,height=\textwidth]{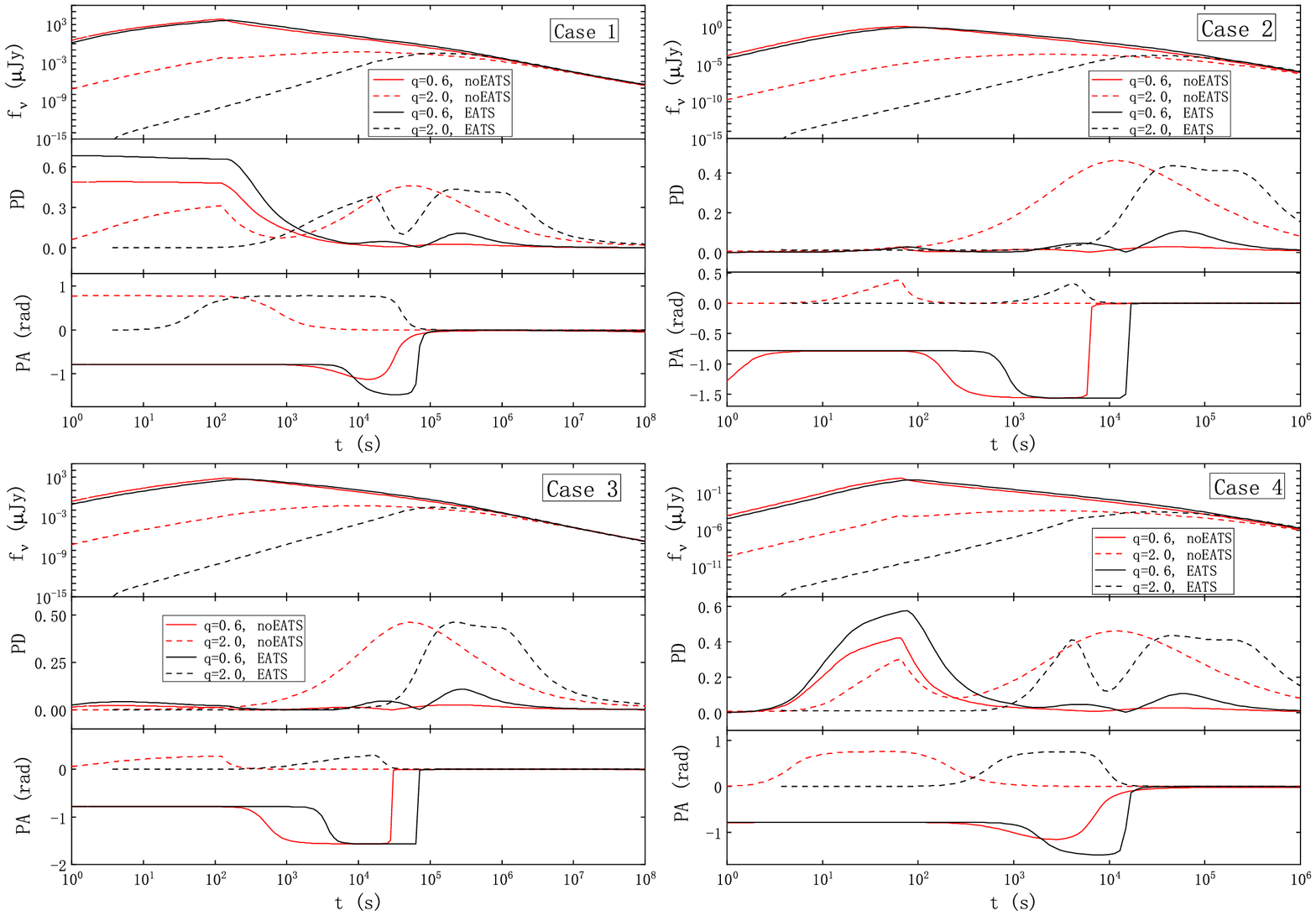}
	\caption{Light curves (upper panel), PD curves (middle panel), and PA curves (lower panel) in four cases with an aligned magnetic field in Region 3. The red lines correspond to old results in \cite{Lan2016} but with corrected polarization treatment for two-dimensional random magnetic field, the black lines represent our new results with EATS effect. The solid and dashed lines correspond to on-axis ($q=0.6$) and off-axis ($q=2.0$) observations, respectively.}
	\label{compare_A}
\end{figure*}

\begin{figure}
	\centering
    \includegraphics[width=\textwidth,height=\textwidth]{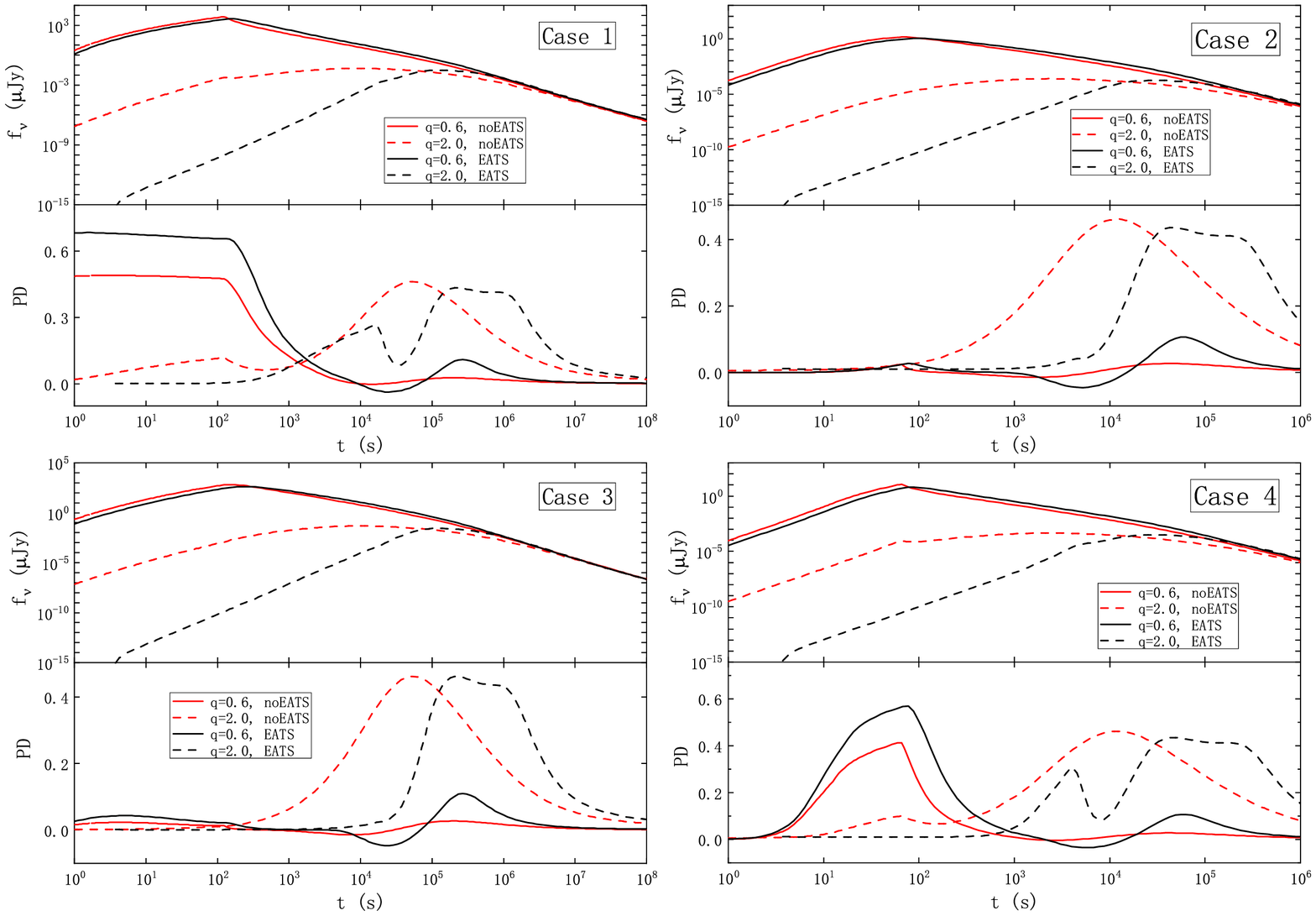}
	\caption{Light curves (upper panel) and PD curves (lower panel) in four cases with a toroidal magnetic field in Region 3. The red lines correspond to old results in \cite{Lan2016} but with corrected polarization treatment for two-dimensional random magnetic field, the black lines represent our new results with the EATS effect. The solid and dashed lines correspond to on-axis ($q=0.6$) and off-axis ($q=2.0$) observations, respectively.}
	\label{compare_T}
\end{figure}

Here, as representations, we consider two observational angles with one on-axis observation ($q\equiv\theta_V/\theta_j=0.6$) and the other off-axis detection ($q=2.0$). Our results are shown in Figs. \ref{compare_A} and \ref{compare_T}. In each case, roughly around the jet break time when $1/\gamma=\theta_V+\theta_j$, there are two small PD bumps for on-axis observation and the PA changes abruptly by $90^\circ$ between the two PD bumps \citep{Sari1999,Rossi2004}, while there is only one large PD bump for large off-axis detection of the forward-shock emission \citep{Rossi2004}. For off-axis detection, PDs of the four cases at late times are all larger than 0 here, while they are negative in \citep{Lan2016} with wrong polarization treatment for the two-dimensional random field. In each case, the light curves and PD curves of an aligned field are similar to that of a toroidal fields for both $q=0.6$ and $q=2.0$. Independent of the observational angle, PA changes (not necessarily $90^\circ$) usually happen around the minimum values of the PD curves \citep{Lan2018,Lan2019PJ}.

For on-axis observation, the profiles of the light curves, PD curves, and PA curves are all similar for the calculations with and without EATS effect. Because of the EATS effect considered and local PD $\pi_0$ (Eq. 1) used here, PD values of our new results are larger than that in \cite{Lan2016}. The PD peaks at late times, mainly due to the forward-shock emission, will be larger with EATS effect for $q<1$, which is consistent with that in \cite{Rossi2004}. While they are comparable and only a temporal shift is expected with EATS effect for $q>1$. With the EATS effect, the peak times of the reverse-shock emission will move to later times for off-axis detection, which is different from a fixed peak time (i.e., independent of $q$) of reverse-shock radiation without EATS effect in \cite{Lan2016}.

For large off-axis detection ($q=2.0$), with the EATS effect, the light curves becomes steeper before their peak times \citep{Huang2007,Rossi2004}, the peak times of both the reverse-shock and the forward-shock emission will shift to later observational times and will increase with $q$, and the peak times of the PA curves with an aligned magnetic field are also delayed. With EATS effect, two peaks in the PD curves of Cases 1 and 4 (the reverse-shock-emission-dominated cases) and the only peaks of the PD curves in Cases 2 and 3 (the forward-shock-emission-dominated cases) all shift to later observational times. The first peaks in the PD curves of Cases 1 and 4 are due to the comparable flux from Region 3 (which is highly polarized) to that from Region 2 \footnote{The peak times of the light curves of the emission from Region 3 are around the first peaks in the PD curves of Cases 1 and 4.}. The second peaks in the PD curves of Cases 1 and 4 and the only peaks of the PD curves in Cases 2 and 3 are mainly because of off-axis observations of forward-shock region, and the peak times will be around the times of the jet break when $1/\gamma=\theta_V+\theta_j$. Because of EATS effect considered, there are two small peaks in the PD bump of Case 2 and 3 with $q=2.0$ at late times compared with the one peak in the PD bump for the non-EATS cases. Because the dynamics for Cases 1 and 3 are same, the peak time of the second PD bump in Case 1 and of the only PD bump in Case 3 are both around the time when $1/\gamma=\theta_V+\theta_j$, hence are almost the same, so does for the peak time of the second PD bump in Case 4 and of the only PD bump in Case 2.

\section{Afterglow polarizations in a stratified medium}
In this section, the parameters of the ejecta are fixed to be: $E_{iso}=10^{50}$ erg, $\eta=100$, and $\Delta_0=3\times10^{10}$ cm. The shocks are also assumed to be adiabatic. The effects of the parameters of the stratified medium ($k$ and $r_0$) on the dynamics are studied. The redshift of the source is fixed to be $z=0.3$.

\begin{figure}
	\centering \includegraphics[width=\textwidth,height=\textwidth]{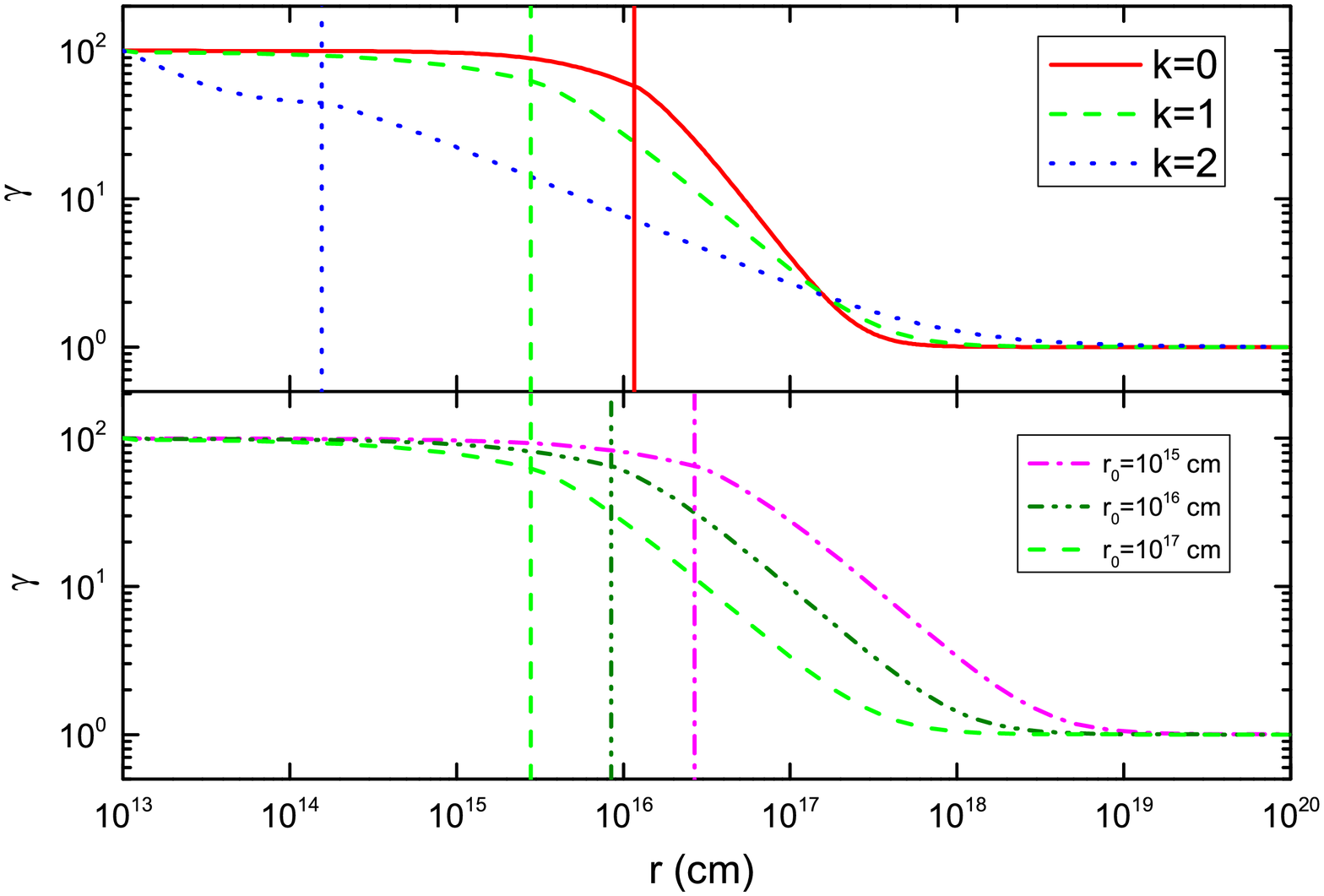}
	\caption{The effects of the parameters of the stratified medium on the dynamics. The upper and lower panels show the dynamics with different $k$ and different $r_0$ values, respectively. In the upper panel, $r_0$ is fixed to be $10^{17}$ cm. The red-solid, green-dashed, and blue-dotted lines correspond to $k=0$, 1, and 2, respectively. In the lower panel, we fix $k=1$. The magenta-dash-dot and olive-dash-dot-dot lines correspond to $r_0=10^{15}$ cm and $10^{16}$ cm, respectively. The vertical lines show the corresponding reverse shock crossing radii.}
	\label{dyn}
\end{figure}

The dynamics used in this section are shown in Fig. \ref{dyn}. In the upper panel, we fix $r_0=10^{17}$ cm and let $k$ to be variable. The number density $n(r)$ with $r<10^{17}$ cm will increase with $k$, leading to stronger shocks with higher velocities. Therefore, the reverse shock crossing radius will decrease with the increase of parameter $k$. In the lower panel, we fix $k=1$ and let $r_0$ to be variable. The number density $n(r)$ at same radius $r$ will be larger for a larger $r_0$, so the shocks will also be stronger for the dynamics with larger $r_0$. Same as the upper panel for the variable $k$ parameter, the reverse shock crossing radius will decrease with the increase of $r_0$.

If the Lorentz factor of Region 3 relative to Region 4 $\gamma_{34}\gg1$, the reverse shock will be ultrarelativistic, corresponding to the thick shell case. And if $\gamma_{34}-1\ll1$, the reverse shock is Newtonian, corresponding to the thin shell case. Sari \& Pian (1995) had discussed the analytical solution of the outflow dynamics and pointed out that if $\eta^2\gg f\equiv n_4/n_1$ the reverse shock is ultrarelativistic and otherwise it is Newtonian. $n_4$ and $n_1$ are the number density of Region 4 and Region 1, respectively. With the calculation, we get  $(f/\eta^2)_{k=2}(r)=(f/\eta^2)_{k=0}(r)(r/r_0)^2$. At the reverse shock crossing radius of $k=2$ case ($R_c=1.56\times10^{14}$ cm), $(f/\eta^2)_{k=0}(R_c)=4.8\times10^4\gg1$ and $(f/\eta^2)_{k=2}(R_c)=(f/\eta^2)_{k=0}(R_c)(R_c/r_0)^2=0.1\ll1$. So at $r=R_c=1.56\times10^{14}$ cm, the reverse shock for $k=0$ is Newtonian while it is relativistic for $k=2$. Since whether or not the reverse shock is relativistic also depends on the parameters of the outer medium (i.e., $k$ and $r_0$), in the following we will not distinguish the thick shell and the thin shell.

In this section, the local PD in an ordered field is presented in Eq. 1, the EATS effect is included, and the polarization treatment for a two-dimensional random field is corrected. In the following, we consider two cases, i.e., the reverse-shock emission dominates the early afterglow (Case I) and the forward-shock radiation dominates the emission during the whole afterglow phase (Case II). We take $\epsilon_{e,3}=\epsilon_{B,3}=\epsilon_{e,2}=0.1$ and $\epsilon_{B,2}=0.01$ for Case I and $\epsilon_{e,3}=\epsilon_{B,3}=10^{-6}$ and $\epsilon_{e,2}=\epsilon_{B,2}=0.1$ for Case II. With the parameters we take for Case II, the ratio $f_{32}\equiv f_{\nu,3}/f_{\nu,2}$ will be smaller than 0.001, so we will neglect the contributions of the reverse-shock region to the total Stokes parameters for Case II. Therefore, with the assumed parameters, there are two emission regions including both the reverse-shock region and the forward-shock region for Case I. And there is finally only one emission region (the forward-shock region) for Case II. As shown in Section 3, PD curves with an aligned field are similar to that with a toroidal field. So in the following we only consider an aligned field in Region 3 and a two-dimensional random field in Region 2. The sets of the parameters in this section are presented in Table \ref{Table 2}.

\begin{table*}[!htbp]
	\caption{Dynamical and Emission Parameters of the Cases I and II}
	\label{Table 2}
	\begin{center}
		\centering
		\begin{tabular}{ccccccccccc}
			\hline\hline\noalign{\smallskip}
			cases  &  $E_{iso}$&$\eta$&$\Delta_0$&$\epsilon_{e,3}$&$\epsilon_{B,3}$&$\epsilon_{e,2}$&$\epsilon_{B,2}$&\\
			\hline\noalign{\smallskip}
		    Case I&$10^{50}$ erg&100&$3\times10^{10}$ cm&0.1&0.1&0.1&0.01&\\[5pt]
		    Case II&$10^{50}$ erg&100&$3\times10^{10}$ cm&$10^{-6}$&$10^{-6}$&0.1&0.1&\\[5pt]
            \noalign{\smallskip}
			\hline
		\end{tabular}
	\end{center}
    \footnotesize{\ \ \ \ \ \ \ \ \ \ \ \ \ \ \ \ \ \ \ \ \ \ \ \ The other parameters are the same for two cases: $\theta_j=0.1$ rad, $p_2=p_3=2.5$, $n_0=1$ cm$^{-3}$, and $z=0.3$.}
\end{table*}

\subsection{Multi-band}
\begin{figure}
	\centering \includegraphics[width=\textwidth,height=\textwidth]{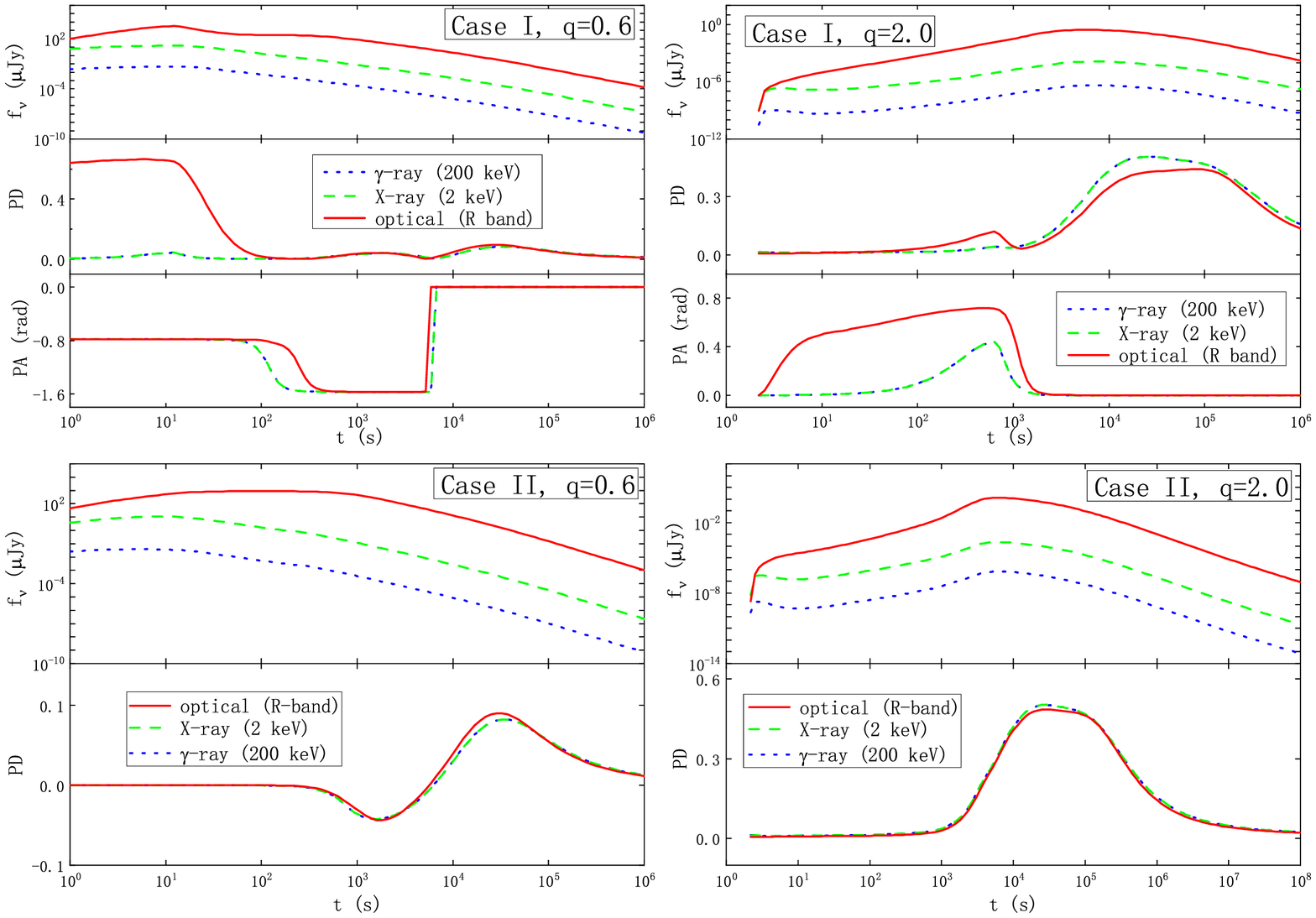}
	\caption{Light curves (upper panel), PD curves (middle panel), and PA curves (lower panel) for different observational energy band of Case I are shown in the first row. Light curves (upper panel) and PD curves (middle panel) for various observational energy band of Case II are presented in the second row. The red-solid, green-dashed, and blue-dotted lines correspond to the observational energy band of optical R-band, X-ray (2 keV), and $\gamma$-ray (200 keV), respectively.}
	\label{multiband}
\end{figure}

With our improved model, we give the polarization predictions of the multi-band GRB afterglows, including wavelength of optical (R-band), X-ray (2 keV), and $\gamma-$ray (200 keV). The parameters of the stratified medium are set to be the typical values of $k=1$ and $r_0=10^{17}$ cm \citep{Yi2013}, and the dynamics used in this subsection is shown as green-dashed lines in Fig. 3. The results of the polarization evolutions are shown in Fig. \ref{multiband}.

Independent of different observational angles and of different Cases, polarization curves (including both the PD and PA curves) almost coincide for the observational energy band of 2 keV and 200 keV. For on-axis observations in Case I, the reverse shock emission, which is highly polarized, dominates the total flux density during the early reverse-shock crossing stage at optical R-band ($f_{32}\sim20\gg1$), so there is a high-level PD plateau before reverse-shock crossing time. The small PD peaks in both X-ray and $\gamma$-ray bands at reverse-shock crossing time are due to the small amount flux contributions of highly polarized emission from Region 3 ($f_{32}\sim0.01\ll1$). Therefore, PD evolution before reverse-shock crossing time is mainly determined by $f_{32}$ for on-axis observations. For off-axis detection in Case I, the first PD peaks in three observational energy band are due to the highly polarized emission from Region 3. The second PD peaks are because of the off-axis observations of the forward shock emission. The differences of the PD evolutions between three observational energy band are tiny.

For the on-axis and off-axis observations in Case II, PD curves are almost coincide for the three observational energy band. Same as in Section 3, if the forward-shock emission dominates the whole afterglow phase, there are two small PD bumps for on-axis observation and there is only one large PD bump for large off-axis detection. For on-axis observation, the first PD bump begins around 200s, roughly correspond to $1/\gamma=\theta_j-\theta_V$. The second PD bump begins around $10^4$ s, roughly correspond to the jet break time at $1/\gamma=\theta_j+\theta_V$. For off-axis detection, the only PD bump begins around $10^3$ s (roughly correspond to $1/\gamma=\theta_V-\theta_j$) and reaches it peak around $2\times10^4$ s (roughly correspond to $1/\gamma=\theta_j+\theta_V$). Therefore, PD bumps in Case II (i.e., the forward-shock emission dominates the whole afterglow radiation) are due to the evolutions of the bulk Lorentz factor.

\subsection{Various observational angles}
\begin{figure}
	\centering \includegraphics[width=\textwidth,height=\textwidth]{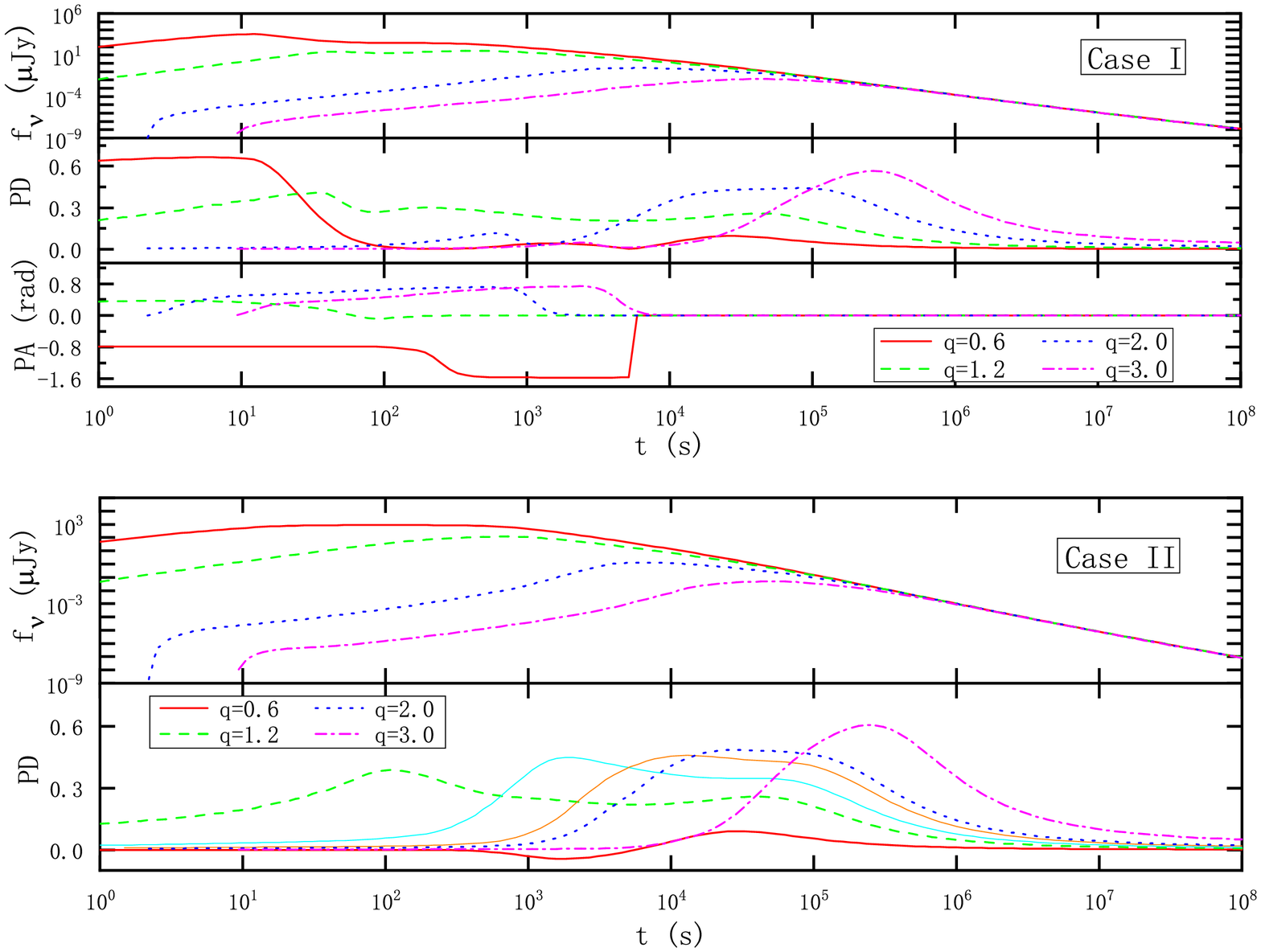}
	\caption{Light curves (upper panel), PD curves (middle panel), and PA curves (lower panel) for different q value of Case I are shown in the first row. Light curves (upper panel) and PD curves (middle panel) for various q value of Case II are presented in the second row. The red-solid, green-dashed, blue-dotted, and magenta-dash-dot lines correspond to $q=0.6$, 1.2, 2.0, and 3.0, respectively. For illustration, the blue and orange solid lines in the second row for Case II correspond to $q=1.5$ and 1.8, respectively.}
	\label{q}
\end{figure}

Observational geometry also affects polarization properties of the jet emission significantly \citep{Waxman2003}. The light curves and polarization properties will be similar for on-axis observations (i.e., $q\leq1$) of an aligned field \citep{Lan2016}. And the EATS effect on the light curves and on the polarization curves become important for off-axis detections, as shown in Section 3. Therefore, we consider four observational angles, one on-axis observations with $q=0.6$ and three off-axis detections with $q=1.2$, 2.0, and 3.0. The results are shown in Fig. \ref{q}. In this subsection, we set the parameters of the stratified medium as their typical value $k=1$ and $r_0=10^{17}$ cm \citep{Yi2013} and the dynamics is shown as green-dashed line in Fig. \ref{dyn}. The observational frequency is set at optical R-band.

The peak times of the light curves from the reverse-shock region will increase with $q$, so does for the forward-shock region. Depending on the ratio $f_{32}$ and on the decrease of bulk Lorentz factor $\gamma$, the situation is very complicated for slightly off-axis detection $q=1.2$. In Case I, a relatively large $f_{32}$ leads to a relatively high PDs ($\sim30\%$) at early times. The decrease of the bulk Lorentz factor will lead to two effects, corresponding to two PD peaks at late observational times. One PD peak will appear around observational time when $1/\gamma=\theta_V-\theta_j$ \citep{Waxman2003} and the other will happen around the jet break time when $1/\gamma=\theta_j+\theta_V$.

In our calculation, dynamics are the same for Cases I and II. Because the forward-shock radiation will dominate the total flux during the whole afterglow phase in Case II, to study the effect of the decaying bulk Lorentz factor on the polarization properties, we focus our analysis on Case II. For $q=1.2$, the first PD bump begins to rise around 10 s, just around the observational time when $1/\gamma=\theta_V-\theta_j$ (corresponding to $\gamma\sim50$ around 16 s). The second PD bump peaks around $4.2\times10^4$ s, after the jet break time when $1/\gamma=\theta_j+\theta_V$ (corresponding to $\gamma\sim4.5$ around $1.9\times10^4$ s).

For large off-axis observational angles (i.e., $q=2.0$ and 3.0) in Case II, The times of the only PD peaks are always slightly after the peak times of the corresponding light curves. The PD peaks will rise around the time when $1/\gamma=\theta_V-\theta_j$ and reach their maximum value around the time when $1/\gamma=\theta_j+\theta_V$ for both $q=2.0$ and 3.0. Because $\theta_V$ will be larger for $q=3.0$ compared with $q=2.0$, the $\gamma$ value at $1/\gamma=\theta_j+\theta_V$ will be smaller for $q=3.0$. So the PD peaks around $1/\gamma=\theta_j+\theta_V$ will shift to late observational times with the increase of $q$.

In Case II, for slight off-axis observations (i.e., the forward-shock-dominated case), there will be two PD bumps in the PD curve, While there will be only one PD bump for large off-axis observations. The first and second PD bumps of the slight off-axis observations begin around $1/\gamma=\theta_V-\theta_j$ and $1/\gamma=\theta_j+\theta_V$, respectively. So the difference of the bulk Lorentz factor between the two beginning times of the PD bumps reads $\Delta\gamma=2/\theta_j/(q^2-1)$. With the increase of the observational angle $q$, $\Delta\gamma$ will decrease. Because the dynamics for Case II is independent of $q$ values, the time interval between the two PD bumps will shorter with the increase of $q$. So when the beginning time of the second PD bump is smaller than the end time of the first PD bump, the two PD bumps of the slight off-axis observations will begin to merge to one PD bump. With the calculation, the convergence of the two PD bumps into one will happen when $1.5<q<1.8$ under the dynamics used here, and the trend is consistent with the results shown in Fig. 8 of \cite{Rossi2004}. This is consistent with the statement reached just above that the only one PD bump also begins around $1/\gamma=\theta_V-\theta_j$ and reaches its peak around $1/\gamma=\theta_j+\theta_V$ for large off-axis observations.

\cite{GK2003} had also considered the PD evolutions for $q<1$ with a two-dimensional random magnetic field in GRB afterglow phase. However, PD signs of our results are opposite with their results, but consistent with that in \cite{Sari1999} and \cite{GL1999}. For large q with $q>1$, the only PD bump in PD curve will move to late observational time and its peak value will increase with an increasement of q value. This trend is consistent with the case for a two-dimensional random magnetic field in the emission region of \cite{Rossi2004} and \cite{Pedreira2022}.

\subsection{The effects of the $k$ parameter}
\begin{figure}
	\centering \includegraphics[width=\textwidth,height=\textwidth]{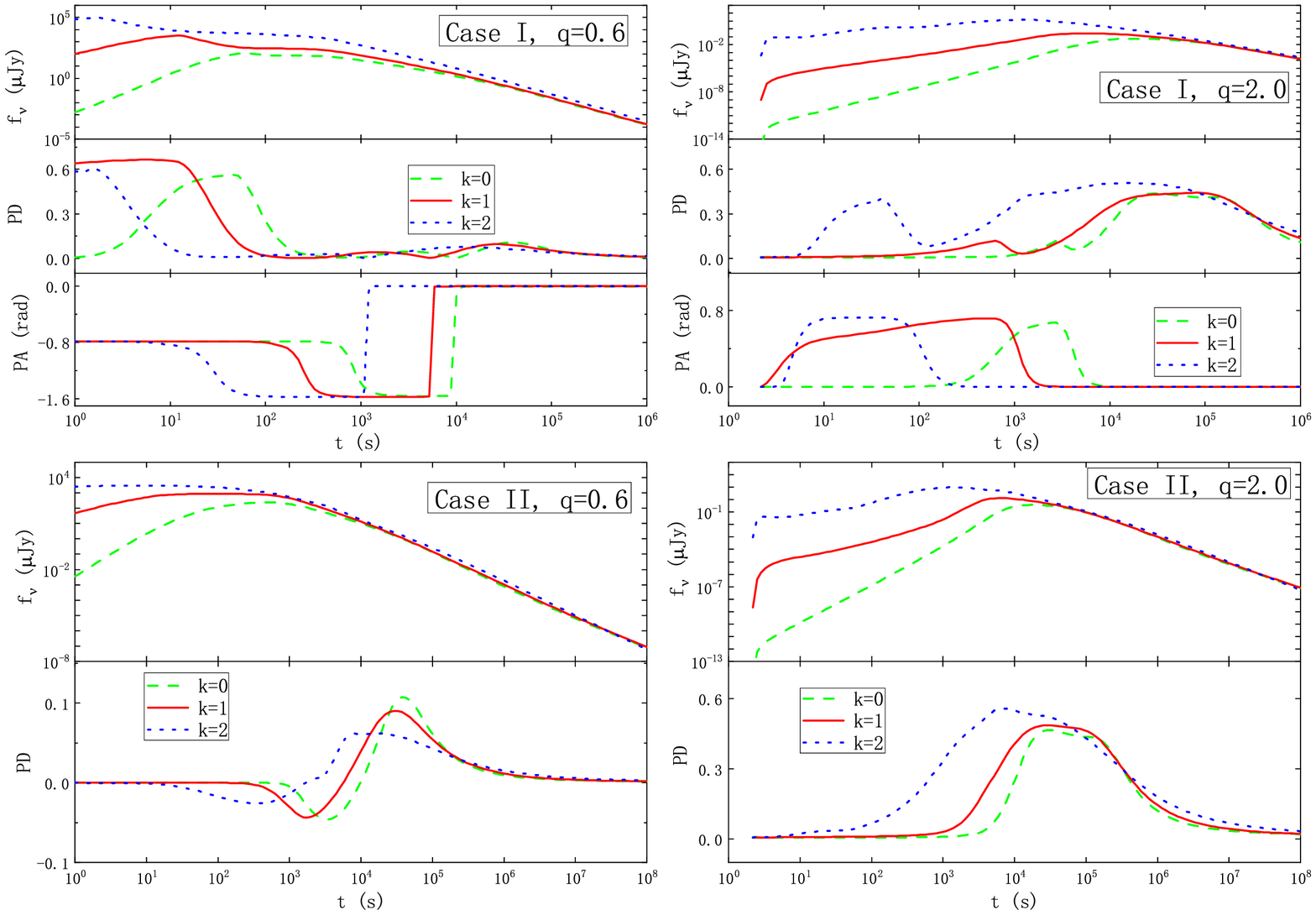}
	\caption{Light curves (upper panel), PD curves (middle panel), and PA curves (lower panel) for different k value of Case I are shown in the first row. Light curves (upper panel) and PD curves (middle panel) for various k value of Case II are presented in the second row. The green-dashed, red-solid, and blue-dotted lines represent $k=0$, 1, and 2, respectively.}
	\label{k}
\end{figure}

In this subsection, we investigate the effects of the $k$ parameter on the light curves and on the polarization properties. We fix $r_0=10^{17}$ cm and the observational frequency is set at optical R-band. The dynamics for $k=0$, 1, and 2 are shown as red-solid, green-dashed, and blue-dotted lines in Fig. \ref{dyn}. The results are shown in Fig. \ref{k}.

Since we take $n_0=1$ cm$^{-3}$ and $r_0=10^{17}$ cm, a larger $k$ will lead to a larger number density of the outer medium when $r<r_0$ and then lead to stronger shocks. Therefore, the flux density at early observational times will increase with $k$ value. Because the reverse-shock crossing time becomes shorter with a stronger reverse shock, the peaks of the light curves in Cases I (the reverse-shock emission dominates the early radiation) will shift to early times with a larger $k$ value.

For on-axis observation ($q=0.6$) of Case I, depending on the ratio $f_{32}$, the early PD curve can be a plateau (for $k=1$ and 2) or a bump (for $k=0$). The early emission in Case I with $q=0.6$ is dominated by reverse-shock region, we calculate the PD of the emission from Region 3 and it reads
\begin{equation}
PD_3\equiv\frac{\sqrt{Q_{\nu,3}^2+U_{\nu,3}^2}}{f_{\nu,3}}
\end{equation}
We find that for each $k$ value initially there is a PD plateau of $PD_3$ curve before the reverse-shock crossing time, then the PD curve will decrease, finally it will increase to a roughly constant value of 0.77 at late observational times. The values of $PD_3$s at their plateau phases are 0.705 (for $k=0$), 0.686 (for $k=1$), and $\sim0.650$ (for $k=2$), respectively. $PD_3$ at plateau phase and $k$ are negatively correlated. From Fig. 3, the bulk Lorentz factor decays faster with a larger $k$ before the reverse-shock crossing radius, which means at same radius the bulk Lorentz factor will be smaller for larger $k$ value (i.e., a larger $1/\gamma$ cone with a larger $k$ value). Because the cancellation effect of polarization over a larger $1/\gamma$ cone becomes more important, $PD_3$ will be smaller with a larger $1/\gamma$ cone. Therefore, the values of $PD_3$ at their early plateau phases are negatively correlated with $k$ value.

To interpret the evolution of the $PD_3$ curve, we calculate the $\tilde{f}_3$ parameter and it is defined as follows \citep{Lan2020}
\begin{equation}
\tilde{f_3}\equiv\frac{\int_{\theta_{min}}^{1/\gamma}df_{\nu,3}}{\int_{1/\gamma}^{\theta_{max}}df_{\nu,3}}
\end{equation}
where $\theta_{min}=0$ for on-axis observations and $\theta_{min}=\theta_V-\theta_j$ for off-axis detections, while $\theta_{max}=min(\theta_V+\theta_j,\theta(r_b))$ and $\theta(r_b)$ corresponds to the $\theta$ value at $r_b$ on one EATS. In Fig. 7, we find $PD_3$ and $\tilde{f}_3$ parameter are positively correlated for $q=0.6$, hence the evolutions of the $PD_3$ curves are mainly determined by the $\tilde{f}_3$ parameter for on-axis observations. For off-axis detection $q=2.0$, after $t\sim10^4$ s (after the time when $1/\gamma=\theta_V-\theta_j$), $PD_3$ is positively correlated to $\tilde{f}_3$, while before $t\sim10^4$ s, $\tilde{f}_3$ equals to 0, however the $PD_3$ still evolve with time. The reason for the $PD_3$ evolution when $\tilde{f}_3=0$ may be because of the changes of the asymmetry of the system with the evolution of $\gamma$, however, we do not have a proper parameter to depict this. Finally, $\tilde{f}_3$ will reach infinity (totally low-latitude emission) after $3.6\times10^4$ s for $q=0.6$ and after $2.5\times10^5$ s for $q=2.0$.

\begin{figure}
	\centering \includegraphics[width=\textwidth,height=\textwidth]{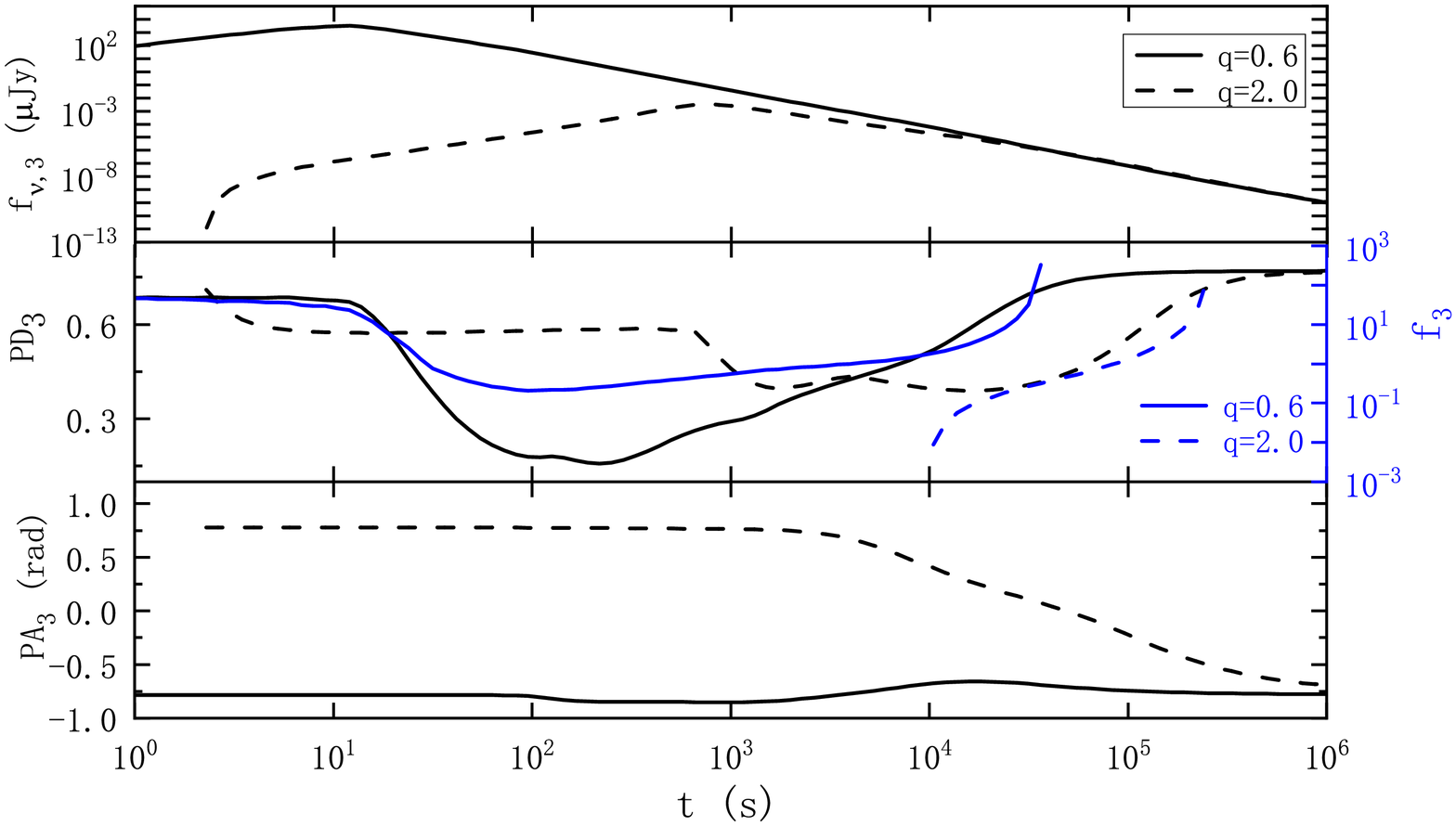}
	\caption{Light curves (upper panel), PD curves (middle panel), and PA curves (lower panel) of the reverse-shock radiation in Case I with $q=0.6$ (solid lines) and 2.0 (dashed-lines). And also the evolutions of the $\tilde{f}_3$ parameter (blue lines) are shown with respect to the right axis of the middle panel.}
	\label{PD3}
\end{figure}

For off-axis detection ($q=2.0$), $PA_3$ is initially positive with a value of 0.7774 rad, and then changes gradually to a negative value of -0.6949 rad with a change of $\Delta PA_3\sim\pi/2$. At the beginning of the evolution, $\tilde{f}_3=0$ (It means that all the emissions are from the high-latitude region without $1/\gamma$ cone.), while at late stage, it is $\infty$ (It means that all the emissions come from low-latitude region within $1/\gamma$ cone.). So the PAs of the low- and high-latitude emissions will have a roughly $90^\circ$ difference.  $PD_3$ is roughly 0.77 for the low-latitude emission (with $\tilde{f}_3=\infty$). Its value for the high-latitude emission (with $\tilde{f}_3=0$) is about 0.6 and is relatively high, which is very different from a low PD value of the high-latitude emission during the GRB prompt phase \citep{Lan2020,Lan2021}.

For Case I with $q=2.0$, the first PD bumps for different $k$ values are around the peak times of the $f_{\nu,3}$ curves, corresponding to the reverse-shock crossing times. Because the shocks will become stronger with an increasing $k$ value, the crossing time of the reverse shock becomes shorter, thus the positions of the first PD bumps will decrease with k. The second PD bump is because of the off-axis detection of the forward-shock radiation. For the same $1/\gamma$ value, the corresponding $r$ or $t$ will be smaller for larger $k$. Therefore, the second PD bump will begin and peak at early observational time for larger $k$ value.

For Case II with $q=0.6$ and 2.0, because the bulk Lorentz factor $\gamma$ decreases faster before $r<10^{17}$ cm for larger $k$ value, leading to smaller observational times both when $1/\gamma=\theta_j-\theta_V$ and when $1/\gamma=\theta_j+\theta_V$ for larger k. Therefore, both the two PD bumps for $q=0.6$ and the only one PD bump for $q=2.0$ shift toward short observational times with an increasing $k$ value. With the increasement of k, the polarization evolution is slower for both $q=0.6$ and 2.0. And our results for the wind environment with $k=2$ are consistent with that shown in Fig. 1 of \cite{Lazzati2004}.

\subsection{The effects of the normalization parameter $r_0$}

\begin{figure}
	\centering \includegraphics[width=\textwidth,height=\textwidth]{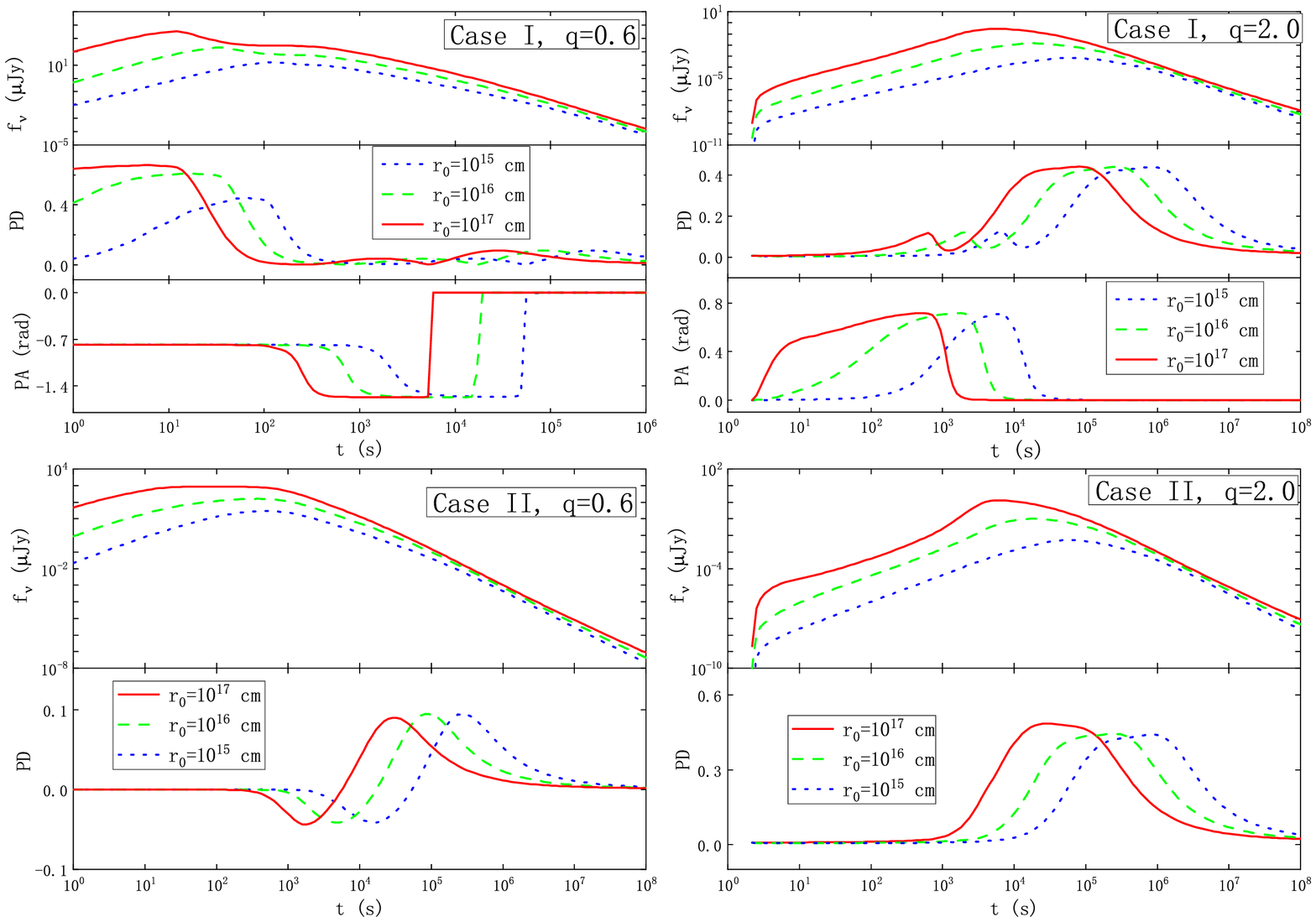}
	\caption{Light curves (upper panel), PD curves (middle panel), and PA curves (lower panel) for different $r_0$ of Case I shown in the first row. Light curves (upper panel) and PD curves (middle panel) for various $r_0$ of Case II shown in the second row. The blue-dotted, green-dashed, and red-solid lines represent $r_0=10^{15}$, $10^{16}$, and $10^{17}$ cm, respectively.}
	\label{r0}
\end{figure}

The effects of the normalization parameter $r_0$ on the polarization evolutions are also considered and the results are shown in Fig. \ref{r0}. We take $k=1$ and the observational frequency is set at the optical R-band. Three values of $r_0$s are considered (i.e., $10^{15}$ cm, $10^{16}$ cm, and $10^{17}$ cm). Dynamics for $r_0=10^{15}$ cm, $10^{16}$ cm, and $10^{17}$ cm are shown as magenta-dash-dot, olive-dash-dot-dot, and green-dashed lines in the lower panel of Fig. 3. The number density $n(r)$ with a larger $r_0$ at radius $r$ will be larger, which will lead to stronger shocks (corresponding to shorter reverse-shock crossing time), then to a higher flux density. Therefore, in Case I the flux density will increase and the peak time of the light curve will decrease with $r_0$ for on-axis observations.

For on-axis observation ($q=0.6$) of Case I, same as that for different $k$ values in Section 4.3, depending on the ratio $f_{32}$, PD curves can be a bump ($r_0=10^{15}$ cm) or a plateau ($r_0=10^{17}$ cm) at early times. For off-axis observation ($q=2.0$) of Case I, a larger $r_0$ will also lead to a stronger shock then to a shorter reverse-shock crossing time, so the first PD bumps (corresponding to the peak of the emission from Region 3) with larger $r_0$ will shift to early observational times. A larger $r_0$ will lead to a smaller bulk Lorentz factor $\gamma(r)$, so for the same $1/\gamma$, $r$ (or the corresponding observational time $t$) will be smaller for a larger $r_0$. Therefore, the second PD bumps (due to the off-axis observation of the forward-shock emission) also shift to early observational time with an increasing $r_0$.

Because the bulk Lorentz factor $\gamma$ decreases earlier for larger $r_0$ value (see the lower panel of Fig. 3), leading to smaller observational times both when $1/\gamma=\theta_j-\theta_V$ and when $1/\gamma=\theta_j+\theta_V$ for larger $r_0$. Therefore, both the two PD bumps for $q=0.6$ and the only PD bumps for $q=2.0$ shift toward short observational time with the increasing $r_0$ value.

\section{Conclusions and Discussion}
In this paper, we consider the EATS effect to study the light curves and polarization evolutions of the GRB afterglows. We assume a large-scale ordered aligned field in the reverse-shock region, so our PD results at early stage give the upper limit. We compare our new results with that in \cite{Lan2016} without EATS effect. Then, we apply our model to predict the multi-band afterglow polarizations in a stratified medium with a power-law number density distribution. And the effects of the observational angle and of the parameters of the stratified medium on the afterglow polarizations are also discussed.

The dynamics in a stratified medium, including both the reverse-shock and forward-shock regions, are studied. We found that for the fixed $n_0$ and $r_0$ value, the bulk Lorentz factor for larger $k$ value will decrease faster at early stage and then decrease slower at late stage. For the fixed $n_0$ and $k$ value, the bulk Lorentz factor for larger $r_0$ value will decrease earlier. The reverse-shock crossing radius will shift to smaller radius with the increase of $k$ or $r_0$.

We found the EATS effect on the afterglow polarizations becomes important for off-axis observations. For the forward-shock-dominated case, there will be only one large PD bump for large off-axis detection, and this PD bump begins roughly at $1/\gamma=\theta_V-\theta_j$ and peaks roughly around the jet break time when $1/\gamma=\theta_j+\theta_V$. Compared with the non-EATS case, the amplitude of the PD bump is similar with the EATS effect, but the peak will shift to late observational time \cite{Rossi2004}. For the reverse-shock-dominated case, there will be two PD bumps for off-axis detection, the PD value of the first PD bump is determined by $f_{32}$, while its value at the second PD bump is usually determined by both the $q$ value and the evolution of the bulk Lorentz factor $\gamma$. Compared with a negative PD values in \cite{Lan2016} for off-axis detections at late observational times, PDs will be larger than 0 with the corrected polarization treatment for two-dimensional random magnetic field in the forward-shock region.

For on-axis observations, assuming the large-scale ordered field in the outflow carried from a central engine, PD value at early afterglow phase is mainly determined by the value of $f_{32}$, i.e., the larger the $f_{32}$, the higher the PD of the jet emission, and vice versa. When $f_{32}\gg1$, PD of the jet emission will reach its maximum value of $PD_3$. When $f_{32}\ll1$, PD will be roughly 0. There are two small PD bumps around the jet break time due to the forward-shock emission \citep{Sari1999}. The first and second PD bumps begin around $1/\gamma=\theta_j-\theta_V$ and $1/\gamma=\theta_V+\theta_j$ (the jet break time), respectively. And there will be an abrupt $90^\circ$ PA change between the two PD bumps \citep{Sari1999}.

We found given the magnetic field configurations in the emission regions, the jet structure, and the observational angles, the evolutions of the $PD_3$ with an ordered magnetic field in Region 3 is mainly determined by $\tilde{f}_3$ parameter (positively correlated), while the evolutions of the afterglow polarizations of the whole jet emission are mainly determined by both the value of $f_{32}$ or the evolutions of bulk Lorentz factor $\gamma$. Through the influences on $f_{32}$ or $\gamma$, various observational energy band, observational angles, $k$ values, and $r_0$ values will finally affect the evolutions of the PD curves.

\acknowledgments
This paper is dedicated to the 70th anniversary of the physics of Jilin University.
This work is supported by the National Natural Science Foundation of China (grant Nos. 11903014, 11833003, and 12041306), the National Key Research and Development Program of China (grant nos. 2022SKA0130100 and 2017YFA0402600), the National SKA Program of China No. 2020SKA0120300, International Partnership Program of Chinese Academy of Sciences for Grand Challenges (114332KYSB20210018), and the Major Science and Technology Project of Qinghai Province (2019-ZJ-A10).

\bibliography{rsfs_eats}

\begin{thebibliography}{}
\expandafter\ifx\csname natexlab\endcsname\relax\def\natexlab#1{#1}\fi

\bibitem[{{Abbott} {et~al.}(2017){Abbott}, {Abbott}, {Abbott}, {Acernese},
  {Ackley}, \& {(INTEGRAL}}]{Abbott2017}
{Abbott}, B.~P., {Abbott}, R., {Abbott}, T.~D., {et~al.} 2017, \apjl, 848, L13

\bibitem[{{Blandford} \& {Znajek}(1977)}]{BZ1977}
{Blandford}, R.~D., \& {Znajek}, R.~L. 1977, \mnras, 179, 433

\bibitem[{{Chevalier} \& {Li}(2000)}]{CL2000}
{Chevalier}, R.~A., \& {Li}, Z.-Y. 2000, \apj, 536, 195

\bibitem[{{Drenkhahn}(2002)}]{Drenkhahn2002}
{Drenkhahn}, G. 2002, \aap, 387, 714

\bibitem[{{Ghisellini} \& {Lazzati}(1999)}]{GL1999}
{Ghisellini}, G., \& {Lazzati}, D. 1999, \mnras, 309, L7

\bibitem[{{Granot} \& {K{\"o}nigl}(2003)}]{GK2003}
{Granot}, J., \& {K{\"o}nigl}, A. 2003, \apjl, 594, L83

\bibitem[{{Gruzinov}(1999)}]{Gruzinov1999}
{Gruzinov}, A. 1999, \apjl, 525, L29

\bibitem[{{Huang} {et~al.}(2007){Huang}, {Lu}, {Wong}, \& {Cheng}}]{Huang2007}
{Huang}, Y.-F., {Lu}, Y., {Wong}, A. Y.~L., \& {Cheng}, K.~S. 2007, \cjaa, 7,
  397

\bibitem[{{Kobayashi}(2000)}]{Kobayashi2000}
{Kobayashi}, S. 2000, \apj, 545, 807

\bibitem[{{Lan} \& {Dai}(2020)}]{Lan2020}
{Lan}, M.-X., \& {Dai}, Z.-G. 2020, \apj, 892, 141

\bibitem[{{Lan} {et~al.}(2019{\natexlab{a}}){Lan}, {Geng}, {Wu}, \&
  {Dai}}]{Lan2019}
{Lan}, M.-X., {Geng}, J.-J., {Wu}, X.-F., \& {Dai}, Z.-G. 2019{\natexlab{a}},
  \apj, 870, 96

\bibitem[{{Lan} {et~al.}(2021){Lan}, {Wang}, {Xu}, {Liu}, \& {Wu}}]{Lan2021}
{Lan}, M.-X., {Wang}, H.-B., {Xu}, S., {Liu}, S., \& {Wu}, X.-F. 2021, \apj,
  909, 184

\bibitem[{{Lan} {et~al.}(2016){Lan}, {Wu}, \& {Dai}}]{Lan2016}
{Lan}, M.-X., {Wu}, X.-F., \& {Dai}, Z.-G. 2016, \apj, 816, 73

\bibitem[{{Lan} {et~al.}(2018){Lan}, {Wu}, \& {Dai}}]{Lan2018}
---. 2018, \apj, 860, 44

\bibitem[{{Lan} {et~al.}(2019{\natexlab{b}}){Lan}, {Xue}, {Xiong}, {Lei}, {Wu},
  \& {Dai}}]{Lan2019PJ}
{Lan}, M.-X., {Xue}, R., {Xiong}, D., {et~al.} 2019{\natexlab{b}}, \apj, 878,
  140

\bibitem[{{Lazzati} {et~al.}(2004){Lazzati}, {Covino}, {Gorosabel}, {Rossi},
  {Ghisellini}, {Rol}, {Castro Cer{\'o}n}, {Castro-Tirado}, {Della Valle}, {di
  Serego Alighieri}, {Fruchter}, {Fynbo}, {Goldoni}, {Hjorth}, {Israel},
  {Kaper}, {Kawai}, {Le Floc'h}, {Malesani}, {Masetti}, {Mazzali}, {Mirabel},
  {Moller}, {Ortolani}, {Palazzi}, {Pian}, {Rhoads}, {Ricker}, {Salmonson},
  {Stella}, {Tagliaferri}, {Tanvir}, {van den Heuvel}, {Wijers}, \&
  {Zerbi}}]{Lazzati2004}
{Lazzati}, D., {Covino}, S., {Gorosabel}, J., {et~al.} 2004, \aap, 422, 121

\bibitem[{{Mazzali} {et~al.}(2003){Mazzali}, {Deng}, {Tominaga}, {Maeda},
  {Nomoto}, {Matheson}, {Kawabata}, {Stanek}, \& {Garnavich}}]{Mazzali2003}
{Mazzali}, P.~A., {Deng}, J., {Tominaga}, N., {et~al.} 2003, \apjl, 599, L95

\bibitem[{{Mundell} {et~al.}(2013){Mundell}, {Kopa{\v{c}}}, {Arnold}, {Steele},
  {Gomboc}, {Kobayashi}, {Harrison}, {Smith}, {Guidorzi}, {Virgili},
  {Melandri}, \& {Japelj}}]{Mundell2013}
{Mundell}, C.~G., {Kopa{\v{c}}}, D., {Arnold}, D.~M., {et~al.} 2013, \nat, 504,
  119

\bibitem[{{Pedreira} {et~al.}(2022){Pedreira}, {Fraija}, {Dichiara}, {Veres},
  {Dainotti}, {Galvan-Gamez}, {Becerra}, \& {Betancourt
  Kamenetskaia}}]{Pedreira2022}
{Pedreira}, A.~C. C. d. E.~S., {Fraija}, N., {Dichiara}, S., {et~al.} 2022,
  arXiv e-prints, arXiv:2210.12904

\bibitem[{{Rossi} {et~al.}(2004){Rossi}, {Lazzati}, {Salmonson}, \&
  {Ghisellini}}]{Rossi2004}
{Rossi}, E.~M., {Lazzati}, D., {Salmonson}, J.~D., \& {Ghisellini}, G. 2004,
  \mnras, 354, 86

\bibitem[{{Sari}(1998)}]{Sari1998}
{Sari}, R. 1998, \apjl, 494, L49

\bibitem[{{Sari}(1999)}]{Sari1999}
---. 1999, \apjl, 524, L43

\bibitem[{{Spruit} {et~al.}(2001){Spruit}, {Daigne}, \&
  {Drenkhahn}}]{Spruit2001}
{Spruit}, H.~C., {Daigne}, F., \& {Drenkhahn}, G. 2001, \aap, 369, 694

\bibitem[{{Steele} {et~al.}(2009){Steele}, {Mundell}, {Smith}, {Kobayashi}, \&
  {Guidorzi}}]{Steele2009}
{Steele}, I.~A., {Mundell}, C.~G., {Smith}, R.~J., {Kobayashi}, S., \&
  {Guidorzi}, C. 2009, \nat, 462, 767

\bibitem[{{Waxman}(2003)}]{Waxman2003}
{Waxman}, E. 2003, \nat, 423, 388

\bibitem[{{Wu} {et~al.}(2003){Wu}, {Dai}, {Huang}, \& {Lu}}]{Wu2003}
{Wu}, X.~F., {Dai}, Z.~G., {Huang}, Y.~F., \& {Lu}, T. 2003, \mnras, 342, 1131

\bibitem[{{Wu} {et~al.}(2005){Wu}, {Dai}, {Huang}, \& {Lu}}]{Wu2005}
---. 2005, \mnras, 357, 1197

\bibitem[{{Yi} {et~al.}(2013){Yi}, {Wu}, \& {Dai}}]{Yi2013}
{Yi}, S.-X., {Wu}, X.-F., \& {Dai}, Z.-G. 2013, \apj, 776, 120

\bibitem[{{Zhang} \& {Kobayashi}(2005)}]{ZK2005}
{Zhang}, B., \& {Kobayashi}, S. 2005, \apj, 628, 315

\end{thebibliography}

\end{document}